__________________________________________________________________________
\documentstyle[12pt]{article}
\title{Correlations of spin states\\
for icosahedral double group} 

\author{Shi-Hai Dong$^{1)}$, Xi-Wen Hou$^{1,2)}$, and Zhong-Qi Ma$^{1)}$\\
{\footnotesize  1) Institute of High Energy Physics, P.O.Box 918(4), Beijing 
100039, P. R. of China}\\
{\footnotesize  2) Department of Physics, University of Three Gorges, 
Yichang 443000, P. R. of China}}

\date{}
\begin{document}
\maketitle

\begin{abstract}
The irreducible bases of the group space of the
icosahedral double groups {\bf I'} and {\bf I$_{h}'$}
are calculated explicitly. Applying those bases on the 
spin states $|j,\mu \rangle$, we present a simple
formula to combine the spin states into the symmetrical 
adapted bases, belonging to a given row of a given irreducible 
representations of {\bf I'} and {\bf I$_{h}'$}.
\end{abstract}

\vspace{10mm}
\noindent
{\bf 1. Introduction}

\vspace{3mm}
Metallo-fullerenes are fullerene cages with a metal atom 
or atoms in the center of the cage. Since Heath et al. 
\cite{hea} showed that metal-containing fullerenes can 
be generated, metallo-fullerenes have drawn considerable 
attentions of physicists and chemists. In order to classify 
the electronic states of such Metallo-fullerenes in
the present of spin-orbit coupling, especially for the 
electronic states with half-odd-integer spin, one has to
deal with the double group symmetry \cite{be}. 

Recently, the character table and the correlation tables 
relevant for the icosahedral group {\bf I$_{h}'$} were 
presented apparently \cite{bal}. As is well known, the correlation 
tables is calculated from the character table by the
standard method of group theory \cite{ham}. From the correlation
table the states with a low angular momentum can be combined 
by a similarity transformation into a state belonging to a given 
row of an irreducible representation of {\bf I'}. However, it 
becomes a tedious task while the angular momentum increases. 
Fortunately, the difficulty can be conquered by the irreducible 
bases in the group space of {\bf I'}. In this note we present
a simple formula (21) to combine the spin states into the symmetrical 
adapted bases, belonging to a given row of a given irreducible 
representations of {\bf I'} and {\bf I$_{h}'$}. The irreducible
bases in the group space of ${\bf I'}$ and the combinations
of the spin states are useful for the calculations of the
vibrational and rotational problems for the carbon-60
molecule \cite{cy}.

From group theory (\cite{ham} p.106), a group element $R$
plays a role of a basis in the group space, which is
the representation space of the regular representation.
The number of times each irreducible representation 
contained in the regular representation is equal to the 
dimension of the representation. Reducing the regular 
representation, we obtain the new bases $\psi^{\Gamma}_{\mu \nu}$
belonging to the $\mu$ ($\nu$) row of the irreducible 
representation $\Gamma$ in the left-(right-)action of
a group element:
$$R\psi^{\Gamma}_{\mu \nu}=\displaystyle \sum_{\rho}~
\psi^{\Gamma}_{\rho \nu}D^{\Gamma}_{\rho \mu}(R),~~~~~~
\psi^{\Gamma}_{\mu \nu}R=\displaystyle \sum_{\rho}~
D^{\Gamma}_{\nu \rho}(R)\psi^{\Gamma}_{\mu \rho}. 
 \eqno (1) $$

\noindent
$\psi^{\Gamma}_{\mu \nu}$ are called the irreducible bases
in the group space. Assume that $G$ is a point group, which is
a subgroup of the rotation group $SO(3)$. Applying the irreducible
bases to the angular momentum states $|j, \rho \rangle$, 
we obtain the combinations $\psi^{\Gamma}_{\mu \nu}~|j, \rho \rangle$,
if it is not vanishing, belonging to the $\mu$ row of the 
representation $\Gamma$ of the point group:
$$R\psi^{\Gamma}_{\mu \nu}~|j, \rho \rangle=\displaystyle \sum_{\tau}~
D^{\lambda}_{\tau \mu}(R)\psi^{\lambda}_{\tau \nu}|j, \rho \rangle.
\eqno (2) $$

\noindent
This method is effective for both integer and half-odd-integer
angular momentum states. In this note we will calculate the
irreducible bases in the group space of the icosahedral
double group (Sec. 2 and 3), and then, find a simple and unified 
formula ((21) in Sec.4) for calculating the combinations. A 
simple conclusion is given in Sec. 5.

Recently, we read a preprint \cite{chen} where 
a technique, called the double-induced technique, was used for
calculating the irreducible bases for the tetrahedral group 
and the combinations of the angular momentum states.
From the preprint we know that a similar work for the 
icosahedral double group is in preparation.

\vspace{4mm}
\noindent
{\bf 2. Icosahedral double group}

\vspace{3mm}
A regular icosahedron is shown in Fig.1. The vertices on the upper
part are labeled by $A_{j}$, $0\leq j \leq 5$, and their opposite 
vertices by $B_{j}$. The $z$ and $y$ axes point from the center
$O$ to $A_{0}$ and the midpoint of $A_{2}B_{5}$, respectively.

\begin{center}
\fbox{Fig. 1.}
\end{center}

The group {\bf I} has 6 five-fold axes, 10 three-fold axes, and 15 
two-fold axes. One of the five-fold axes directs along $z$ axis, and
the rest point from $B_{j}$ to $A_{j}$ ($1 \leq j \leq 5$) with the
polar angle $\theta_{1}$ and azimuthal angles $\varphi_{j}^{(1)}$. 
The rotations through $2\pi/5$ around those five-fold axes are denoted by 
$T_{j}$, $0\leq j \leq 5$. The three-fold axes join the centers of two 
opposite faces. The polar angles of the first and last 5 axes 
are $\theta_{2}$ and $\theta_{3}$, respectively, and 
the azimuthal angles $\varphi_{j}^{(2)}$.  The rotations
through $2\pi/3$ around those three-fold axes are denoted by $R_{j}$, 
$1\leq j \leq 10$. The two-fold axes join the midpoints of two 
opposite edges. The polar and azimuthal angles of the first, next 
and last 5 axes are $\theta_{4}$, $\varphi_{j}^{(1)}$, $\theta_{5}$, 
$\varphi_{j}^{(2)}$, $\pi$, and $\varphi_{j}^{(3)}$, respectively.
The rotations through $\pi$ around those two-fold axes are denoted 
by $S_{j}$, $1 \leq j \leq 15$. Those angles $\theta_{i}$ and 
$\varphi_{j}^{(i)}$ are given as follows:
$$\begin{array}{lll}
\tan \theta_{1}=2,~~  &\tan \theta_{2}=3-\sqrt{5},~~ 
&\tan \theta_{3}=3+\sqrt{5},\\
\tan \theta_{4}=\left(\sqrt{5}-1\right)/2,~~
&\tan \theta_{5}=\left(\sqrt{5}+1\right)/2,~~& \\
\varphi_{j}^{(1)}=2(j-1)\pi/5, &\varphi_{j}^{(2)}=(2j-1)\pi/5,
&\varphi_{j}^{(3)}=(4j-3)\pi/10 .
\end{array} \eqno (3) $$

As is well known, $SU(2)$ group is the covering group of
the rotation group $SO(3)$, and provides the double-valued
representations of $SO(3)$. In order to classify the
angular momentum states with half-odd-integer spin, we
have to extend the point group to the double point group,
following the homomorphism of $SU(2)$ onto $SO(3)$:
$$\pm u(\hat{{\bf n}},\omega)~\longrightarrow ~R(\hat{{\bf n}},\omega).
\eqno (4) $$

In the rotation group $SO(3)$, a rotation through $2\pi$
is equal to identity $E$, but it is different from identity
in the $SU(2)$ group:
$$R(\hat{{\bf n}},2\pi)=E,~~~~~u(\hat{{\bf n}},2\pi)
\equiv E'=-{\bf 1}.  \eqno (5) $$

\noindent
Similarly, a point group $G$ is extended into a double point
group $G'$ by introducing a new element $E'$, satisfying:
$$RE'=E'R,~~~~~(E')^{2}=E,~~~~~R\in G\subset G',~~~~E'R\in G'.
\eqno (6) $$

\noindent
The point group $G$ is a subgroup of $SO(3)$, and the double point
group $G'$ is that of $SU(2)$. For definiteness, we restrict the 
rotation angle $\omega$ not larger than $\pi$:
$$\begin{array}{l}
R(\hat{{\bf n}},\omega)~\longrightarrow~u(\hat{{\bf n}},\omega),\\
R(\hat{{\bf n}},\omega-2\pi)=R(-\hat{{\bf n}},2\pi-\omega)~
\longrightarrow~u(-\hat{{\bf n}},2\pi-\omega)
=-u(\hat{{\bf n}},\omega),\end{array}
~~~~~0\leq \omega \leq \pi. \eqno (7) $$

\noindent
The period of $\omega$ in $SU(2)$ group is $4\pi$. The element
$E'$ was denoted by $R$ in \cite{bal} and \cite{ham}, and by 
$\theta$ in \cite{chen}. The double point group $G'$ was
denoted by $G^{\dagger}$ in \cite{chen}.

The icosahedral double group {\bf I'} contains 120 elements and
nine classes. There are nine inequivant irreducible representations
for {\bf I'}: Five representations $A$, $T_{1}$, $T_{2}$, $G$
and $H$ are called single-valued ones, and four representations
$E_{1}'$, $E_{2}'$, $G'$ and $I'$ are double-valued ones. The row
(column) index runs over integer (in a single-valued representation)
or half-odd-integer (in a double-valued one) as follows:
$$\begin{array}{llll}
A:~~&m=0,~~~~~~~~~& E_{1}':~~&\mu=1/2,~ -1/2, \\
T_{1}: &m=1,~0,~-1, &E_{2}': &\mu=3/2,~ -3/2, \\
T_{2}: &m=2,~0,~-2, &G': &\mu=3/2,~ 1/2,~-1/2,~-3/2, \\
G: &m=2,~1,~-1,~-2,~~ &I': &\mu=5/2,~3/2,~1/2,~-1/2,~-3/2,~-5/2, \\
H: &m=2,~1,~0,~-1,~-2, &
\end{array} \eqno (8) $$

\noindent
where, as in the angular momentum theory, the subscript $\mu$
is replaced by $m$ when it is integer.

The icosahedral double group {\bf I$_{h}'$} is the direct product of
{\bf I'} and the inversion group $\{E,P\}$, where $P$ is the 
inversion operator. According to the parity, the irreducible 
representations of {\bf I$_{h}'$} are denoted as $\Gamma_{g}$
(even) and $\Gamma_{u}$ (odd), respectively. The character table
of the double group {\bf I$_{h}'$} was listed in Table 1 of
\cite{bal}. In this note we will pay more attention to
the double group {\bf I'}. 

\vspace{4mm}
\noindent
{\bf 3. Irreducible bases}

\vspace{3mm}
The rank of the double group {\bf I'} is three. We choose 
$T_{0}$, $S_{1}$ and $E'$ as the generators of {\bf I'}. 
The representation matrix of $E'$ is equal to the unit
matrix {\bf 1} in a single-valued irreducible representation
and {\bf -1} in a double-valued one.
It is convenient to choose the bases in an irreducible representations
of {\bf I'} such that the representation matrices of the 
generator $T_{0}$ are diagonal with the diagonal elements $\eta^{\mu}$. 
Assume that the bases $\Phi_{\mu \nu}$ in the {\bf I'} group space are the 
eigenstates of left-action and right-action of $T_{0}$:
$$\begin{array}{ll}
T_{0}~\Phi_{\mu \nu}=\eta^{\mu}\Phi_{\mu \nu},~~~~~
&\Phi_{\mu \nu}~T_{0}=\eta^{\nu}\Phi_{\mu \nu}, \\
\end{array} \eqno (9) $$

\noindent
where the constant $\eta$ satisfies the following equations:
$$\begin{array}{l}
\eta=\exp(-i2\pi/5),~~~~~\displaystyle \sum_{m=0}^{4}~\eta^{m}=0, \\ 
p=\eta+\eta^{-1}=(\sqrt{5}-1)/2,~~~~
p^{-1}=1+\eta+\eta^{-1}=(\sqrt{5}+1)/2,\\
q=i\left(\eta-\eta^{-1}\right)=\left(\sqrt{5}p^{-1}\right)^{1/2},~~~~
i\left(\eta^{2}-\eta^{-2}\right)=qp.
\end{array} \eqno (10) $$

The bases $\Phi_{\mu \nu}$ can be easily calculated by the projection 
operator $P_{\mu}$ (see p.113 in \cite{ham}): 
$$\Phi_{\mu \nu}=c~P_{\mu}~R~P_{\nu},~~~~~
P_{\mu}=\displaystyle {1 \over 10} \sum_{a=0}^{4}~
\eta^{-\mu a}~\left(E+\eta^{-5\mu}E'\right)
T_{0}^{a}, \eqno (11) $$

\noindent
where $c$ is a normalization factor. The choice of the
group element $R$ in (11) will not affect the results except for
the factor $c$. The subscripts $\mu$ and $\nu$ should be
integer or half-odd-integer, simultaneously. In the following 
we choose $E$, $S_{11}$, $S_{5}$ and $S_{10}$ as the group 
element $R$, respectively, and obtain four independent sets 
of bases $\Phi_{\mu \nu}^{(i)}$: 
$$\begin{array}{rl}
\Phi^{(1)}_{\mu \mu}&=\displaystyle {E+\eta^{-5\mu}E' \over \sqrt{10}} 
\sum_{a=0}^{4}~\eta^{-\mu a}~T_{0}^{a} ,\\
\Phi^{(2)}_{\mu \overline{\mu}}&=\displaystyle 
{E+\eta^{-5\mu}E' \over \sqrt{10}} 
\sum_{a=0}^{4}~\eta^{-\mu a}~T_{0}^{a}S_{11}\\
&=\displaystyle {E+\eta^{-5\mu}E' \over \sqrt{10}} 
\left(S_{11}+\eta^{-2 \mu}S_{12}+\eta^{-4\mu} S_{13} 
+\eta^{4\mu}S_{14}+\eta^{2 \mu}S_{15}\right), \\
\end{array} $$
$$\begin{array}{rl}
\Phi^{(3)}_{\mu \nu}&=\displaystyle {E+\eta^{-5\mu}E' \over 5\sqrt{2}} 
\sum_{a=0}^{4}~\eta^{-\mu a}~T_{0}^{a}S_{5} \sum_{b=0}^{4}~
\eta^{-\nu b}~T_{0}^{b}\\
&=\displaystyle {E+\eta^{-5\mu}E' \over 5\sqrt{2}}
\left\{\left(S_{5}+\eta^{-\mu}R^{2}_{5}
+\eta^{-2 \mu}T^{4}_{1}+\eta^{2 \mu}T_{4}+\eta^{\mu} R_{4} \right)\right.\\
&~~~+~\eta^{(\mu-\nu)}\left(S_{4}+\eta^{-\mu}R^{2}_{4}
+\eta^{-2 \mu}T^{4}_{5}+\eta^{2 \mu}T_{3}+\eta^{\mu} R_{3} \right)\\
&~~~+~\eta^{2(\mu-\nu)}\left(S_{3}+\eta^{-\mu}R^{2}_{3}
+\eta^{-2 \mu}T^{4}_{4}+\eta^{2 \mu}T_{2}+\eta^{\mu} R_{2} \right)\\
&~~~+~\eta^{-2(\mu-\nu)}\left(S_{2}+\eta^{-\mu}R^{2}_{2}
+\eta^{-2 \mu}T^{4}_{3}+\eta^{2 \mu}T_{1}+\eta^{\mu} R_{1} \right)\\
&~~~+~\eta^{-(\mu-\nu)}\left.\left(S_{1}+\eta^{-\mu}R^{2}_{1}
+\eta^{-2 \mu}T^{4}_{2}+\eta^{2 \mu}T_{5}+\eta^{\mu} R_{5} \right)
\right\}~, \\
\Phi^{(4)}_{\mu \nu}&=\displaystyle {E+\eta^{-5\mu}E' \over 5\sqrt{2}} 
\sum_{a=0}^{4}~\eta^{-\mu a}~T_{0}^{a}S_{10} \sum_{b=0}^{4}~
\eta^{-\nu b}~T_{0}^{b}\\
&=\displaystyle {E+\eta^{-5\mu}E' \over 5\sqrt{2}}\left\{\left(S_{10}+\eta^{-\mu}T^{3}_{1}
+\eta^{-2 \mu}R^{2}_{6}+\eta^{2 \mu}R_{9}+\eta^{\mu} T^{2}_{5} \right)\right.\\
&~~~+~\eta^{(\mu-\nu)}\left(S_{9}+\eta^{-\mu}T^{3}_{5}
+\eta^{-2 \mu}R^{2}_{10}+\eta^{2 \mu}R_{8}+\eta^{\mu} T^{2}_{4} \right)\\
&~~~+~\eta^{2(\mu-\nu)}\left(S_{8}+\eta^{-\mu}T^{3}_{4}
+\eta^{-2 \mu}R^{2}_{9}+\eta^{2 \mu}R_{7}+\eta^{\mu} T^{2}_{3} \right)\\
&~~~+~\eta^{-2(\mu-\nu)}\left(S_{7}+\eta^{-\mu}T^{3}_{3}
+\eta^{-2 \mu}R^{2}_{8}+\eta^{2 \mu}R_{6}+\eta^{\mu} T^{2}_{2} \right)\\
&~~~+~\eta^{-(\mu-\nu)}\left.\left(S_{6}+\eta^{-\mu}T^{3}_{2}
+\eta^{-2 \mu}R^{2}_{7}+\eta^{2 \mu}R_{10}+\eta^{\mu} T^{2}_{1} \right)
\right\}~. \end{array} \eqno (12) $$

\noindent
where and hereafter the subscript $\overline{\mu}$ denotes $-\mu$. 
Those bases $\Phi_{\mu \nu}^{(i)}$ should
be combined into the irreducible bases $\psi_{\mu \nu}^{\Gamma}$ 
belonging to the given irreducible representation $\Gamma$. The 
combinations can be determined from the condition that the 
irreducible basis should be the eigenstate of a class operator $W$, 
which was called CSCO-I in \cite{chen}. The eigenvalues 
$\alpha_{\Gamma}$ can be calculated from the characters \cite{bal} 
in the irreducible representations $\Gamma$:
$$\begin{array}{ll}
W=\displaystyle \sum_{j=0}^{5}~\left(T_{j}+E'T_{j}^{4}\right),~~
&W~\psi_{\mu \nu}^{\Gamma}=\psi_{\mu \nu}^{\Gamma}~W=
\alpha_{\Gamma}~\psi_{\mu \nu}^{\Gamma}, \\
\alpha_{A}=12,~~~~\alpha_{T_{1}}=4p^{-1},~~~~
&\alpha_{T_{2}}=-4p,~~~~\alpha_{G}=-3,~~~~
\alpha_{H}=0, \\
\alpha_{E_{1}'}=6p^{-1},~~~~\alpha_{E_{2}'}=-6p,~~~~
&\alpha_{G'}=3,~~~~\alpha_{I'}=-2.
\end{array} \eqno (13) $$

\noindent
Now we calculate the matrix form of $W$ in the bases 
$\Phi_{\mu \nu}^{(i)}$, and diagonalize it. $\psi_{\mu \nu}^{\Gamma}$
are just the eigenvectors of the matrix form of $W$:
$$\psi_{\mu \nu}^{\Gamma}=N^{-1/2}~\displaystyle \sum_{i=1}^{4}~
c_{i}~\Phi_{\mu \nu}^{(i)} , \eqno (14) $$

\noindent
where $N$ is the normalization factor. In principle, 
$\psi_{\mu \nu}^{\Gamma}$ can change a phase depending on
$\mu$ and $\nu$. We choose the phases such that the representation
matrices of {\bf I} coincide with those in the subduced 
representations of $D^{j}$ of SO(3):
$$\begin{array}{lll}
D^{0}(R)=D^{A}(R),~~&D^{1}(R)=D^{T_{1}}(R),~~&D^{2}(R)=D^{H}(R), \\
D^{1/2}(R)=D^{E_{1}'}(R),~~&D^{3/2}(R)=D^{G'}(R),~~
&D^{5/2}(R)=D^{I'}(R).  \end{array} \eqno (15) $$

\noindent
The representation matrices of $E'$ and $T_{0}$ are diagonal with the 
diagonal elements $\pm 1$ and $\eta^{\mu}$, respectively (see (9) ), 
and those of another generator $S_{1}$ of {\bf I} are as follows:

{\small $$\begin{array}{c}
D^{A}(S_{1})=1,~~~
D^{T_{1}}(S_{1})=\displaystyle {1 \over \sqrt{5} }
\left( \begin{array}{ccc} -p^{-1} & -\sqrt{2} & -p \\
-\sqrt{2} & 1 & \sqrt{2} \\ -p & \sqrt{2} & -p^{-1} \end{array} \right),\\
D^{T_{2}}(S_{1})=\displaystyle {1 \over \sqrt{5} }
\left( \begin{array}{ccc} -p & \sqrt{2} & p^{-1}\\
\sqrt{2} & -1 & \sqrt{2} \\ p^{-1} & \sqrt{2} & -p \end{array} \right) ,~~
D^{G}(S_{1})=\displaystyle {1 \over \sqrt{5} }
\left( \begin{array}{cccc} -1 & -p & -p^{-1} & 1 \\
-p & 1 & -1 & -p^{-1} \\ -p^{-1} & -1 & 1 & -p \\
1 & -p^{-1} & -p & -1 \end{array} \right),\\
D^{H}(S_{1})=\displaystyle {1 \over 5 }
\left( \begin{array}{ccccc} p^{-2} & 2p^{-1} & \sqrt{6} & 2p 
& p^{2} \\
2p^{-1} & p^{2} & -\sqrt{6} & -p^{-2} & -2p \\
 \sqrt{6} & -\sqrt{6} & -1 & \sqrt{6} & \sqrt{6} \\ 
2p & -p^{-2} & \sqrt{6} & p^{2} & -2p^{-1} \\ 
p^{2} & -2p & \sqrt{6} & -2p^{-1} & p^{-2} \end{array} \right) ,\\
D^{E_{1}'}(S_{1})=\displaystyle {iq \over \sqrt{5} }
\left( \begin{array}{cc} -1 & -p \\ -p & 1 \end{array} \right),~~~~
D^{E_{2}'}(S_{1})=\displaystyle {iq \over \sqrt{5} }
\left( \begin{array}{cc} -p & -1 \\ -1 & p \end{array} \right), \\
D^{G'}(S_{1})=\displaystyle {iq \over 5 }
\left( \begin{array}{cccc} p^{-1} & \sqrt{3} & \sqrt{3} p & p^{2} \\
\sqrt{3} & -p^{2} & -p^{-1} & -\sqrt{3} p \\ 
\sqrt{3} p & -p^{-1} & p^{2} & \sqrt{3} \\
p^{2} & -\sqrt{3} p & \sqrt{3} & -p^{-1} \end{array} \right),  \\
D^{I'}(S_{1})=\displaystyle {iq \over 5\sqrt{5} }
\left( \begin{array}{cccccc} -p^{-2} & -\sqrt{5} p^{-1} & -\sqrt{10} 
& -\sqrt{10} p & -\sqrt{5} p^{2} & -p^{3} \\
-\sqrt{5} p^{-1} & -\sqrt{5} p & \sqrt{10} p & \sqrt{10} & \sqrt{5} 
& \sqrt{5} p^{2} \\
-\sqrt{10} & \sqrt{10} p & \sqrt{5} & -\sqrt{5} p & -\sqrt{10} 
& -\sqrt{10} p \\ 
-\sqrt{10} p & \sqrt{10} & -\sqrt{5} p & -\sqrt{5} & \sqrt{10} p & \sqrt{10}\\
-\sqrt{5} p^{2} & \sqrt{5} & -\sqrt{10} & \sqrt{10} p & \sqrt{5} p
& -\sqrt{5} p^{-1} \\
-p^{3} & \sqrt{5} p^{2} & -\sqrt{10} p & \sqrt{10} & -\sqrt{5} p^{-1}
& p^{-2} \end{array} \right) .  
\end{array} \eqno (16) $$ }

\noindent
The normalization factors $N$ and combination
coefficients $c_{i}$ are listed in Table 1. 

\begin{center}
\fbox{Table 1}
\end{center}

Now, we obtain the irreducible bases $\psi_{\mu \nu}^{\Gamma}$ 
satisfying (1). The irreducible bases of the group {\bf I$_{h}'$} 
can be expressed as follows:
$$\psi_{\mu \nu}^{\Gamma_{g}}=2^{-1/2} \left(E+P\right)
\psi_{\mu \nu}^{\Gamma},~~~~~~~
\psi_{\mu \nu}^{\Gamma_{u}}=2^{-1/2} \left(E-P\right)
\psi_{\mu \nu}^{\Gamma}. \eqno (17) $$

\vspace{4mm}
\noindent
{\bf 4. Applications to the angular momentum states}

\vspace{3mm}
Due to the properties (1), we can obtain the irreducible
function bases by applying $\psi_{\mu \nu}^{\Gamma}$ to
any function. As an important application, we apply
$\psi_{\mu \nu}^{\Gamma}$ to the angular momentum states
$|j,\mu \rangle$, where the Condon-Shortley definition
is used:
$$R~|j,\mu \rangle=\displaystyle \sum_{\nu=-j}^{j}~
D^{j}_{\nu \mu}(R)~|j, \nu \rangle,~~~~R\in SO(3) {\rm ~or~} SU(2).
\eqno (18) $$ 

\noindent
When $j$ is an integer $\ell$, $|\ell,m\rangle$ is nothing but the 
spherical harmonics $Y^{\ell}_{m}(\theta, \varphi)$. 

From Fig.1. and (3) we have: 
$$\begin{array}{l}
E'~|j, \mu \rangle=(-1)^{2j}~|j, \mu \rangle,~~~~
T_{0}~|j, \mu \rangle=\eta^{\mu}~|j, \mu \rangle,\\
S_{5}~|j,\mu \rangle =\displaystyle \sum_{\nu}~
D^{j}_{\nu \mu}(-2\pi/5, 2\theta_{4}, 7\pi/5)~|j,\nu \rangle
=\displaystyle \sum_{\nu}~
e^{-i\mu \pi}\eta^{\mu-\nu}d^{j}_{\nu \mu}(2\theta_{4})~|j,\nu \rangle \\
S_{10}~|j,\mu \rangle =\displaystyle \sum_{\nu}~
D^{j}_{\nu \mu}(-\pi/5, 2\theta_{5}, 6\pi/5)~|j,\nu \rangle
=\displaystyle \sum_{\nu}~
e^{i\nu \pi}\eta^{3\mu+2\nu}d^{j}_{\nu \mu}(2\theta_{5})~|j,\nu \rangle \\
S_{11}~|j,\mu \rangle =\displaystyle \sum_{\nu}~
D^{j}_{\nu \mu}(0, \pi, 4\pi/5)~|j,\mu \rangle
=(-1)^{j-\mu}\eta^{2\mu}~|j,-\mu \rangle .
\end{array} \eqno (19) $$

\noindent
where $d^{j}(\theta)$ is the usual $D$-function in the angular
momentum theory \cite{ham}, and
$$\cos \theta_{4}=\sin \theta_{5}=q/\sqrt{5},~~~~~
\cos \theta_{5}=\sin \theta_{4}=qp/\sqrt{5}. \eqno (20) $$

Now, it is easy to obtain the combinations of 
the angular momentum states
$\psi_{\mu \lambda}^{\Gamma}~|j,\rho \rangle$, belonging to the 
$\mu$ row of the irreducible representation $\Gamma$ of {\bf I'}:
$$\begin{array}{rl}
\psi_{\mu \lambda}^{\Gamma}~|j,\rho \rangle &=~\sqrt{10/N}
\delta_{\lambda \rho}' \displaystyle \sum_{\nu}~\delta_{\mu \nu}'
~\left\{c_{1}\delta_{\rho \nu}+c_{2}\delta_{\rho \overline{\nu}}
(-1)^{j-\rho} \eta^{2\rho}\right. \\
&\left.~~~+\sqrt{5}c_{3}e^{-i\pi \rho}\eta^{\rho-\nu}
d^{j}_{\nu \rho}(2\theta_{4})
+\sqrt{5}c_{4}e^{i \pi \nu}\eta^{3\rho+2\nu}
d^{j}_{\nu \rho}(2\theta_{5})\right\}~|j, \nu \rangle , 
\end{array} \eqno (21) $$

\noindent
where $N$ and $c_{i}$ were given in Table 1, $\delta_{\lambda \rho}'$ is
defined as follows:
$$\delta_{\lambda \rho}'=\left\{\begin{array}{ll}1~~~~&{\rm when}~~
(\lambda-\rho)/5={\rm integer} \\
0 & {\rm otherwise}. \end{array} \right. \eqno (22) $$

\noindent
In driving (21) some terms were merged so that the functions
need be normalized again.

(21) is our main formula. For fixed $\lambda$ and $\rho$, 
satisfying $\delta_{\lambda \rho}'=1$,
we obtain the combinations of the angular momentum states
$\psi_{\mu \lambda}^{\Gamma}~|j,\rho \rangle$, belonging to the $\mu$
row of the irreducible representation $\Gamma$ of {\bf I'}. 
Different choice of $\lambda$ and $\rho$ may cause the combinations
vanishing, dependent on each other, or independent. The
number of independent combinations depends upon the number of times
that the irreducible representation $\Gamma$ of {\bf I'}
appears in the reduced form of the subduced representation 
of $D^{j}$ of SU(2). The latter is completely determined by the 
character of the representation and listed in Table 2 of \cite{bal}.

Those combinations are very easy to be calculated by a simple 
computer file or even by hand. In the following we list
some combinations as examples.
$$\begin{array}{ll}
\psi_{00}^{A}~|0,0\rangle =2\sqrt{30}~|0,0\rangle,~~~
&\psi_{\mu 1}^{T_{1}}~|1, 1\rangle =2\sqrt{10}~|1,\mu\rangle, \\
\psi_{\mu 2}^{H}~|2, 2\rangle =2\sqrt{6}~|2,\mu\rangle, ~~
&\psi_{\mu (1/2)}^{E_{1}'}~|1/2, 1/2\rangle =-i2\sqrt{15}~|1/2,\mu\rangle, \\
\psi_{\mu (3/2)}^{G'}~|3/2, 3/2\rangle =i\sqrt{30}~|3/2,\mu\rangle,~~ 
&\psi_{\mu (5/2)}^{I'}~|5/2, 5/2\rangle =-i2\sqrt{5}~|5/2,\mu\rangle.
\end{array} $$
$$\begin{array}{l}
\psi_{2\overline{2}}^{T_{2}}~|3, 3\rangle =-4\left(\sqrt{3/5}~|3,2\rangle
+\sqrt{2/5} ~|3,-3 \rangle \right), \\
\psi_{0\overline{2}}^{T_{2}}~|3, 3\rangle =-4~|3,0\rangle, \\
\psi_{\overline{2}\,\overline{2}}^{T_{2}}~|3, 3\rangle =-4\left(
-\sqrt{2/5}~|3,3\rangle +\sqrt{3/5} ~|3,-2 \rangle \right), \\
\psi_{2\overline{2}}^{G}~|3, 3\rangle =3\sqrt{2}\left(
-\sqrt{2/5}~|3,2\rangle +\sqrt{3/5}~ |3,-3 \rangle \right), \\
\psi_{1\overline{2}}^{G}~|3, 3\rangle =3\sqrt{2}~|3,1\rangle, \\
\psi_{\overline{1}\,\overline{2}}^{G}~|3, 3\rangle =3\sqrt{2}~
|3,-1\rangle , \\
\psi_{\overline{2}\,\overline{2}}^{G}~|3, 3\rangle =3\sqrt{2}\left(
\sqrt{3/5}~|3,3\rangle +\sqrt{2/5} ~|3,-2 \rangle \right).
\end{array} $$
$$\begin{array}{l}
\psi_{3/2\,\overline{3/2}}^{E_{2}'}~|7/2, 7/2\rangle =-i3\sqrt{2}
\left(-\sqrt{7/10}~|7/2,3/2\rangle +\sqrt{3/10} ~|7/2,-7/2 
\rangle \right), \\
\psi_{\overline{3/2}\,\overline{3/2}}^{E_{2}'}~|7/2, 7/2\rangle =-i3\sqrt{2}
\left(\sqrt{3/10}~|7/2,7/2\rangle +\sqrt{7/10} ~|7/2,-3/2 
\rangle \right), \\
\psi_{5/2\,\overline{3/2}}^{I'}~|7/2, 7/2\rangle =i\sqrt{14}~
\left(\sqrt{1/50}~|7/2,5/2\rangle+7/\sqrt{50}~|7/2,-5/2 \rangle \right), \\
\psi_{3/2\,\overline{3/2}}^{I'}~|7/2, 7/2\rangle =i\sqrt{14}~
\left(-\sqrt{3/10}~|7/2,3/2\rangle-\sqrt{7/10}~|7/2,-7/2 \rangle \right), \\
\psi_{1/2\,\overline{3/2}}^{I'}~|7/2, 7/2\rangle =i\sqrt{14}~
|7/2,1/2\rangle , \\
\psi_{\overline{1/2}\,\overline{3/2}}^{I'}~|7/2, 7/2\rangle =-i\sqrt{14}~
~|7/2,-1/2 \rangle, \\
\psi_{\overline{3/2}\,\overline{3/2}}^{I'}~|7/2, 7/2\rangle =i\sqrt{14}~
\left(-\sqrt{7/10}~|7/2,7/2\rangle+\sqrt{3/10}~|7/2,-3/2 \rangle \right), \\
\psi_{\overline{5/2}\,\overline{3/2}}^{I'}~|7/2, 7/2\rangle =i\sqrt{14}~
\left(7/\sqrt{50}~|7/2,5/2\rangle-\sqrt{1/50}~|7/2,-5/2 \rangle \right).
\end{array} $$

\vspace{4mm}
\noindent
{\bf 5. Conclusion}

\vspace{3mm}
If the Hamiltonian of a system has a given symmetry, 
the symmetry adapted bases are very useful in calculating
the eigenvalues and eigenstates. From the irreducible bases 
in the group space of the symmetry group of the system, the 
the symmetry adapted bases  can be calculated
generally and simply. The combinations of the angular momentum 
states are important examples for calculating the symmetry 
adapted bases. In this note we calculate the explicit form 
of the irreducible bases of {\bf I'} group space, and obtain
a general formula for calculating the combinations of
angular momentum states into the irreducible basis functions
belonging to the given row of a given irreducible representation
of the icosahedral double group {\bf I'}. This method is
effective for any double point group.

\vspace{5mm}
{\bf Acknowledgments}. The authors would like to thank professor 
Jin-Quan Chen for sending us his preprint \cite{chen} before 
publication. This work was supported by 
the National Natural Science Foundation of China and Grant No. 
LWTZ-1298 of Chinese Academy of Sciences.

\newpage
\begin{center}

{\bf Table 1} $~~$ Irreducible bases in the group space of {\bf I'}
$$\begin{array}{c}
\psi_{\mu \nu}^{\Gamma}=N^{-1/2} \displaystyle \sum_{i=1}^{4}~
c_{i}~\Phi_{\mu \nu}^{(i)},\\
\eta=\exp(-i2\pi/5),~~~~
p=\eta+\eta^{-1},~~~~q=i(\eta-\eta^{-1}). \\[2mm]
\psi_{00}^{A}=\left(\Phi_{00}^{(1)}+\Phi_{00}^{(2)}
+\sqrt{5}\Phi_{00}^{(3)}+\sqrt{5}\Phi_{00}^{(4)}\right)/\sqrt{12} .
\end{array}  $$

{\scriptsize
\begin{tabular}{c|c|ccccc||c|c|ccccc} \hline \hline 
\multicolumn{7}{c||}{$\Gamma=T_{1}$} &
\multicolumn{7}{c}{$\Gamma=T_{2}$} \\ \hline
$\mu$ & $\nu$ & $c_{1}$ & $c_{2}$ & $c_{3}$ & $c_{4}$ & $N$ &
$\mu$ & $\nu$ & $c_{1}$ & $c_{2}$ & $c_{3}$ & $c_{4}$ & $N$ 
\\ \hline

$1$ & $1$ & $1$ &  & $-p^{-1}$ & $-p$ & $4$ &
$2$ & $2$ & $1$ &  & $-p$ & $-p^{-1}$ & $4$ \\
$0$ & $1$ &  &  & $-\eta^{-1}$ & $\eta^{2}$ & $2$ &
$0$ & $2$ &  &  & $\eta^{-2}$ & $-\eta^{-1}$ & $2$ \\
$\overline{1}$ & $1$ & &$\eta^{-2}$ &$-\eta^{-2}p$ &$-\eta^{-1}p^{-1}$ &$4$&
$\overline{2}$ & $2$& & $-\eta$  & $\eta p^{-1}$ & $\eta^{-2} p$ & $4$\\

$1$ & $0$ &  &  & $-\eta$ & $\eta^{-2}$ & $2$ &
$2$ & $0$ &  &  & $\eta^{2}$ & $-\eta$ & $2$ \\
$0$ & $0$ & $1$ & $-1$ & $1$ & $-1$ & $4$ &
$0$ & $0$ & $1$ & $-1$ & $-1$ & $1$ & $4$ \\
$\overline{1}$ & $0$ &  &  & $\eta^{-1}$ & $-\eta^{2}$ &$2$ &
$\overline{2}$ & $0$ &  &  & $\eta^{-2}$ & $-\eta^{-1}$ & $2$\\

$1$ & $\overline{1}$ & & $\eta^{2}$ & $-\eta^{2} p$ & $-\eta p^{-1}$ & $4$ &
$2$ & $\overline{2}$ & &$-\eta^{-1}$ &$\eta^{-1} p^{-1}$ & $\eta^{2} p$ & $4$ \\
$0$ & $\overline{1}$ & & & $\eta$ & $-\eta^{-2}$ & $2$ &
$0$ & $\overline{2}$ & & & $\eta^{2}$ & $-\eta$ & $2$ \\
$\overline{1}$ & $\overline{1}$ & $1$ & & $-p^{-1}$ & $-p$ &$4$ &
$\overline{2}$ & $\overline{2}$ & $1$ &  & $-p$ & $-p^{-1}$ & $4$\\ \hline
\end{tabular} 
}

\vspace{4mm}
{\scriptsize 
\begin{tabular}{c|c|ccccc||c|c|ccccc}  
\multicolumn{14}{c}{$\Gamma=G$} \\ \hline
$\mu$ & $\nu$ & $c_{1}$ & $c_{2}$ & $c_{3}$ & $c_{4}$ & $N$ &
$\mu$ & $\nu$ & $c_{1}$ & $c_{2}$ & $c_{3}$ & $c_{4}$ & $N$ 
\\ \hline

$2$ & $2$ & $1$ &  & $-1$ & $1$ & $3$ &
$2$ & $\overline{1}$ & & &$-\eta^{-2} p^{-1}$ & $-\eta^{-1} p$ & $3$ \\
$1$ & $2$ &  &  & $-\eta^{-1} p$ & $-\eta^{2} p^{-1}$ & $3$ &
$1$ & $\overline{1}$ & & $\eta^{2}$ & $-\eta^{2}$ & $\eta$ & $3$ \\
$\overline{1}$ & $2$ & & &$-\eta^{2} p^{-1}$ &$-\eta p$ &$3$&
$\overline{1}$ & $\overline{1}$ &$1$ & &$1$ &$-1$ &$3$\\
$\overline{2}$ & $2$ & & $\eta$ &$\eta$ &$-\eta^{-2}$ &$3$&
$\overline{2}$ & $\overline{1}$ & & & $-\eta^{-1} p$ & $-\eta^{2} p^{-1}$ 
&$3$\\

$2$ & $1$ &  & & $-\eta p$ & $-\eta^{-2} p^{-1}$ & $3$ &
$2$ & $\overline{2}$ &  & $\eta^{-1}$ & $\eta^{-1}$ & $-\eta^{2}$ & $3$ \\
$1$ & $1$ & $1$ &  & $1$ & $-1$ & $3$ &
$1$ & $\overline{2}$ &  & & $-\eta^{-2} p^{-1}$ & $-\eta^{-1} p$ & $3$ \\
$\overline{1}$ & $1$ & & $\eta^{-2}$ & $-\eta^{-2}$ & $\eta^{-1}$ & $3$ &
$\overline{1}$ & $\overline{2}$ & & &$-\eta p$ &$-\eta^{-2} p^{-1}$ &$3$ \\
$\overline{2}$ & $1$ & & & $-\eta^{2} p^{-1}$ & $-\eta p$ & $3$ &
$\overline{2}$ & $\overline{2}$ &$1$ & &$-1$ &$1$ &$3$ \\ \hline
\end{tabular} 
}

\vspace{4mm}
{\scriptsize 
\begin{tabular}{c|c|ccccc||c|c|ccccc} 
\multicolumn{14}{c}{$\Gamma=H$} \\ \hline
$\mu$ & $\nu$ & $c_{1}$ & $c_{2}$ & $c_{3}$ & $c_{4}$ & $N$ &
$\mu$ & $\nu$ & $c_{1}$ & $c_{2}$ & $c_{3}$ & $c_{4}$ & $N$ 
\\ \hline

$2$ & $2$ & $\sqrt{5}$ &  & $p^{-2}$ & $p^{2}$ & $12$ &
$\overline{1}$ & $0$ & & &$\eta^{-1}$ &$\eta^{2}$ &$2$ \\
$1$ & $2$ &  &  & $\eta^{-1} p^{-1}$ & $-\eta^{2} p$ & $3$ &
$\overline{2}$ & $0$ & & & $\eta^{-2}$ & $\eta^{-1}$ & $2$\\
$0$ & $2$ &  &  & $\eta^{-2}$ & $\eta^{-1}$ & $2$ &
$2$ & $\overline{1}$ & & & $\eta^{-2} p$ & $-\eta^{-1} p^{-1}$ & $3$ \\
$\overline{1}$ & $2$ & & &$\eta^{2} p$ &$-\eta p^{-1}$ &$3$&
$1$ & $\overline{1}$ & & $-\sqrt{5}\eta^{2}$ & $-\eta^{2} p^{-2}$ 
& $-\eta p^{2}$ & $12$ \\
$\overline{2}$ & $2$ & & $\sqrt{5}\eta$ &$\eta p^{2}$ &$\eta^{-2} p^{-2}$ 
&$12$&
$0$ & $\overline{1}$ & &  & $\eta$ & $\eta^{-2}$ & $2$ \\

$2$ & $1$ & & & $\eta p^{-1}$ & $-\eta^{-2} p$ & $3$ &
$\overline{1}$ & $\overline{1}$ & $\sqrt{5}$ & & $p^{2}$ & $p^{-2}$ & $12$\\
$1$ & $1$ & $\sqrt{5}$ & & $p^{2}$ & $p^{-2}$ & $12$ &
$\overline{2}$ & $\overline{1}$ & & & $-\eta^{-1} p^{-1}$ 
& $\eta^{2} p$ & $3$\\
$0$ & $1$ &  & & $-\eta^{-1}$ & $-\eta^{2}$ & $2$ &
$2$ & $\overline{2}$ & & $\sqrt{5}\eta^{-1}$ &$\eta^{-1} p^{2}$ 
& $\eta^{2} p^{-2}$ & $12$ \\
$\overline{1}$ & $1$ & & $-\sqrt{5}\eta^{-2}$ &$-\eta^{-2} p^{-2}$ 
&$-\eta^{-1} p^{2}$ &$12$&
$1$ & $\overline{2}$ & & & $-\eta^{-2} p$ & $\eta^{-1} p^{-1}$ & $3$ \\
$\overline{2}$ & $1$ & & &$-\eta^{2} p$ &$\eta p^{-1}$ &$3$&
$0$ & $\overline{2}$ & & & $\eta^{2}$ & $\eta$ & $2$ \\

$2$ & $0$ &  &  & $\eta^{2}$ & $\eta$ & $2$ &
$\overline{1}$ & $\overline{2}$ & & &$-\eta p^{-1}$ &$\eta^{-2} p$ &$3$\\
$1$ & $0$ &  & & $-\eta$ & $-\eta^{-2}$ & $2$ &
$\overline{2}$ & $\overline{2}$ & $\sqrt{5}$ & & $p^{-2}$ & $p^{2}$ &$12$\\
$0$ & $0$ & $\sqrt{5}$ & $\sqrt{5}$ & $-1$ & $-1$ & $12$ &
&&&&&&\\ \hline 
\end{tabular} 
}
\end{center}

\vspace{4mm}
\begin{center}
{\scriptsize
\begin{tabular}{c|c|ccccc||c|c|ccccc}  \hline 
\multicolumn{7}{c||}{$\Gamma=E_{1}'$} &
\multicolumn{7}{c}{$\Gamma=E_{2}'$} \\ \hline
$2\mu$ & $2\nu$ & $c_{1}$ & $c_{2}$ & $c_{3}$ & $c_{4}$ & $N$ &
$2\mu$ & $2\nu$ & $c_{1}$ & $c_{2}$ & $c_{3}$ & $c_{4}$ & $N$ \\ \hline

$1$ & $1$ & $-i$ &  & $q$ & $qp$ & $6$ &
$3$ & $3$ & $-i$ &  & $qp$ & $-q$ & $6$ \\
$\overline{1}$ & $1$ & & $-i\eta^{-1}$ & $\eta^{-1}qp$ & $-\eta^{2}q$ & $6$ &
$\overline{3}$ & $3$ & & $i\eta^{2}$ & $\eta^{2}q$ & $\eta qp$ & $6$ \\
$1$ & $\overline{1}$ & & $i\eta$ & $\eta qp$ & $-\eta^{-2}q$ & $6$ &
$3$ & $\overline{3}$ & & $-i\eta^{-2}$ & $\eta^{-2} q$ & $\eta^{-1}qp$ & $6$ \\
$\overline{1}$ & $\overline{1}$ & $-i$ & & $-q$ & $-qp$ & $6$ &
$\overline{3}$ & $\overline{3}$ & $-i$ & & $-qp$ & $q$ & $6$ \\ \hline
\end{tabular} 
}

\vspace{4mm}
{\scriptsize 
\begin{tabular}{c|c|ccccc||c|c|ccccc}  
\multicolumn{14}{c}{$\Gamma=G'$} \\ \hline
$2\mu$ & $2\nu$ & $c_{1}$ & $c_{2}$ & $c_{3}$ & $c_{4}$ & $N$ &
$2\mu$ & $2\nu$ & $c_{1}$ & $c_{2}$ & $c_{3}$ & $c_{4}$ & $N$ 
\\ \hline

$3$ & $3$ & $i\sqrt{5}$ &  & $qp^{-1}$ & $qp^{2}$ & $15$ &
$3$ & $\overline{1}$ & & & $\eta^{2} qp$ & $\eta q$ & $5$ \\
$1$ & $3$ &  &  & $\eta^{-1} q$ & $-\eta^{2} qp$ & $5$ &
$1$ & $\overline{1}$ & & $i\sqrt{5}\eta$ & $-\eta qp^{-1}$ 
& $-\eta^{-2}qp^{2}$ & $15$ \\

$\overline{1}$ & $3$ & & &$\eta^{-2} qp$ & $\eta^{-1} q$ & $5$ &
$\overline{1}$ & $\overline{1}$ & $i\sqrt{5}$ & & $qp^{2}$ 
& $-qp^{-1}$ & $15$\\

$\overline{3}$ & $3$ & & $i\sqrt{5} \eta^{2}$ & $\eta^{2} qp^{2}$ 
& $-\eta qp^{-1}$ & $15$ &
$\overline{3}$ & $\overline{1}$ & & & $\eta^{-1} q$ 
& $-\eta^{2} qp$ & $5$ \\

$3$ & $1$ &  & & $\eta q$ & $-\eta^{-2} qp$ & $5$ &
$3$ & $\overline{3}$ &  & $-i\sqrt{5} \eta^{-2}$ & $\eta^{-2}qp^{2}$ 
& $-\eta^{-1}qp^{-1}$ & $15$ \\
$1$ & $1$ & $i\sqrt{5}$ &  & $-qp^{2}$ & $qp^{-1}$ & $15$ &
$1$ & $\overline{3}$ &  & & $-\eta^{2} qp$ & $-\eta q$ & $5$ \\

$\overline{1}$ & $1$ & & $-i\sqrt{5} \eta^{-1}$ & $-\eta^{-1}qp^{-1}$ 
& $-\eta^{2}qp^{2}$ & $15$ &
$\overline{1}$ & $\overline{3}$ & & &$\eta q$ &$-\eta^{-2} qp$ &$5$ \\

$\overline{3}$ & $1$ & & & $-\eta^{-2} qp$ & $-\eta^{-1} q$ & $5$ &
$\overline{3}$ & $\overline{3}$ &$i\sqrt{5}$ & &$-qp^{-1}$ &$-qp^{2}$ &$15$ \\
\hline 
\end{tabular} 
}

\vspace{4mm}
{\scriptsize 
\begin{tabular}{c|c|ccccc||c|c|ccccc} 
\multicolumn{14}{c}{$\Gamma=I'$} \\ \hline
$2\mu$ & $2\nu$ & $c_{1}$ & $c_{2}$ & $c_{3}$ & $c_{4}$ & $N$ &
$2\mu$ & $2\nu$ & $c_{1}$ & $c_{2}$ & $c_{3}$ & $c_{4}$ & $N$ \\ \hline

$5$ & $5$ & $-i5$ &  & $qp^{-2}$ & $qp^{3}$ & $50$ &
$5$ & $\overline{1}$ & & & $\eta^{-2}qp$ & $-\eta^{-1} q$ & $5$ \\
$3$ & $5$ &  &  & $\eta^{-1}qp^{-1}$ & $-\eta^{2} qp^{2}$ & $10$ &
$3$ & $\overline{1}$ & & & $-\eta^{2} q$ & $\eta qp$ & $5$\\
$1$ & $5$ & & & $\eta^{-2} q$ & $\eta^{-1} qp$ & $5$ &
$1$ & $\overline{1}$ & & $i\sqrt{5} \eta$ & $\eta qp$ & $\eta^{-2} q$ & $10$ \\
$\overline{1}$ & $5$ &  &  & $\eta^{2} qp$ & $-\eta q$ & $5$ &
$\overline{1}$ & $\overline{1}$ & $-i\sqrt{5}$ & & $q$ & $-qp$ & $10$\\
$\overline{3}$ & $5$ &  &  & $\eta qp^{2}$ & $\eta^{-2} qp^{-1}$ & $10$ &
$\overline{3}$ & $\overline{1}$ & & & $-\eta^{-1} qp$ & $-\eta^{2} q$ & $5$\\
$\overline{5}$ & $5$ &  & $-i5$ & $qp^{3}$ & $-qp^{-2}$ & $50$ &
$\overline{5}$ & $\overline{1}$ & & & $-\eta^{-2} q$ & $-\eta^{-1} qp$ & $5$\\

$5$ & $3$ & &  & $\eta qp^{1}$ & $\eta^{-2} qp^{2}$ & $10$ &
$5$ & $\overline{3}$ & & & $\eta^{-1} qp^{2}$ & $\eta^{2} qp^{-1}$ & $10$ \\
$3$ & $3$ & $-i\sqrt{5}$ &  & $qp$ & $q$ & $10$ &
$3$ & $\overline{3}$ & & $-i\sqrt{5}\eta^{-2}$ & $-\eta^{-2} q$ 
& $\eta^{-1} qp$ & $10$\\
$1$ & $3$ & & & $-\eta^{-1} qp$ & $-\eta^{2} q$ & $5$ &
$1$ & $\overline{3}$ & & & $\eta^{2} q$ & $-\eta qp$ & $5$ \\
$\overline{1}$ & $3$ &  &  & $-\eta^{-2} q$ & $\eta^{-1} qp$ & $5$ &
$\overline{1}$ & $\overline{3}$ & & & $-\eta qp$ & $-\eta^{-2} q$ & $5$\\
$\overline{3}$ & $3$ &  & $i\sqrt{5}\eta^{2}$ & $-\eta^{2} q$ & $\eta qp$ 
& $10$ &
$\overline{3}$ & $\overline{3}$ & $-i\sqrt{5}$ & & $-qp$ & $-q$ & $10$\\
$\overline{5}$ & $3$ &  &  & $-\eta qp^{2}$ & $-\eta^{-2} qp^{-1}$ & $10$ &
$\overline{5}$ & $\overline{3}$ & & & $\eta^{-1} qp^{-1}$ 
& $-\eta^{2} qp^{2}$ & $10$\\ 

$5$ & $1$ & & & $\eta^{2}q$ & $\eta qp$ & $5$ &
$5$ & $\overline{5}$ & & $i5$ & $qp^{3}$ & $-qp^{-2}$ & $50$ \\
$3$ & $1$ &  &  & $-\eta qp$ & $-\eta^{-2} q$ & $5$ &
$3$ & $\overline{5}$ & & & $-\eta^{-1} qp^{2}$ & $-\eta^{2} qp^{-1}$ & $10$\\
$1$ & $1$ & $-i\sqrt{5}$ & & $-q$ & $qp$ & $10$ &
$1$ & $\overline{5}$ & & & $\eta^{-2} qp$ & $-\eta^{-1} q$ & $5$ \\
$\overline{1}$ & $1$ &  & $-i\sqrt{5}\eta^{-1}$ & $\eta^{-1} qp$ 
& $\eta^{2} q$ & $10$ &
$\overline{1}$ & $\overline{5}$ & & & $-\eta^{2} q$ & $-\eta qp$ & $5$\\
$\overline{3}$ & $1$ &  &  & $\eta^{-2} q$ & $-\eta^{-1} qp$ & $5$ &
$\overline{3}$ & $\overline{5}$ & & & $\eta^{-1} qp^{-1}$ 
& $-\eta^{-2} qp^{2}$ & $10$\\
$\overline{5}$ & $1$ &  & & $\eta^{2} qp$ & $-\eta q$ & $5$ &
$\overline{5}$ & $\overline{5}$ & $-i5$ & & $-qp^{-2}$ & $-qp^{3}$ & $50$\\
\hline \hline
\end{tabular} 
}
\end{center}

\end{document}